 \documentstyle[12pt]{article}
 
\begin{document}
 \titlepage
 \title{
 \begin{flushright}
 Preprint PNPI-1872, May 1993
 \end{flushright}
 \bigskip\bigskip\bigskip
 Explicit calculation of multiloop amplitudes in
 the superstring theory\thanks{This work was supported, in part,
 by a Soros Foundation Grant awarded by the American
 Physical Society.}}
 \date{}
 \author{G.S. Danilov\thanks{E-mail address: danilov@lnpi.spb.su}\\
 Petersburg Nuclear Physics Institute,\\
 Gatchina, 188350, St.-Petersburg, Russia}
 \maketitle
 \begin{abstract}
 Multiloop superstring amplitudes are calculated in the explicit
 form by the solution of Ward identities.  A naive generalization
 of Belavin-Knizhnik theorem to the superstring is found to be
 incorrect since the  period matrix turns out to be depended
 on the spinor structure over the  terms proportional to odd moduli.
 These terms appear because  fermions  mix bosons under the two-dim.
 supersymmetry transformations. The closed, oriented superstring
 turns out  to be finite, if it possesses the ten-dimensional
 supersymmetry, as well  as the two-dimentional one.
 This problem needs a further study.
 \end{abstract}
 \newpage

 \section{Introduction}
 In the well known scheme [1-4]  the superstring
 amplitudes are obtained by summation over the spinning string ones.
 Every spinning string amplitude does not satisfy supersymmetry.
 It turns out to be a source of serious difficulties [3,4]
 in the scheme above.  Recently the manifestly supersymmetrical
 scheme has  been proposed [5,6].  It generalizes the results
 of ref. [7] to the superstring theory.  In the presented paper
 the discussed scheme [5,6] is applied to the explicit calculation of
 the multiloop amplitudes in the closed, oriented superstring theory.
 We consider the even spin structures, the odd spin ones being planned to
 discuss in another paper. Also, we consider only the boson emission
 amplitudes.

 In the considered scheme the superstring amplitudes  are calculated from
 equations that are none  other than Ward identities. These equations
 realize the requirement that  the superstring amplitudes  are
 independent of both the  "vierbein" and the  gravitino field.
 The above equations  determine the partition functions  except only for
 arbitrary constant factors, some number of them  being reduced by the
 supermodular  invariance. To calculate ( in the terms of a coupling
 constant )  all  these factors one should use the  unitarity
 equations. Instead we use the  factorization requirement on the
 superstring amplitudes when  two handles move away from each other.
 So, the superstring amplitudes turn out to be fully
 determined by the gauge  invariance  together with the "factorization
 requirement" above.

 As soon as  fermions mix  bosons under
 the supersymmetry  transformations,  the period matrix  appears to be
 depended on the spinor structure in the terms proportional to odd moduli.
 Because of this effect  the naive  generalization [1,3,4] of the
 Belavin-Knizhnik theorem [8] to the  superstring  is  found to
     be incorrect.

 The problem of the divergences  needs a further study even
  in the closed, oriented  superstring theory. In this theory
 the possible divergences arise when  the handles move away
 from each other.  These divergences disappear, if the known
 "nonrenormalization theorems" [9] are valid. However,
 in the presented paper we do not verify the above theorems
 because of the mathematical complexity of this verification.

 A different approach to the discussed problems has been proposed
 recently  in ref. [10].

 \section{Superspin structures }
 In the  supercovariant scheme [5,6] a  genus-n
 superstring amplitude is  found to be the sum over  "superspin"
 structures integrated over $(3n-3|2n-2)$ complex  moduli. If all the
 odd moduli are taken to be equal to zero, then   every  genus-n
 superspin structure  $(l_1 ,l_2)$  is reduced to  the ordinary
  $(l_1,l_2)$ spin one. Here $l_1$ and $l_2$ are the theta function
  characteristics:  $(l_1 ,l_2)=\bigcup_s(l_{1s} ,l_{2s})$
  where $l_{is}\in(0,1/2)$. The (super)spin structure is even, if
  $l_1l_2=\sum_{s=1}^ nl_{1s}l_{2s}=0$.  It is odd, if $4l_1l_2=1$.

  To every (super)spin structure one can  assign  the "transition"
  group. The above transition groups  are defined on the $( 1|1)$
  complex supermanifolds [11] mapped by the  supercoordinate
  t=$(z|\theta)$; z is a local complex coordinate and $\theta$ is
  its odd partner. The transition group  is generated by its base
  elements ($\Gamma_{a,s}$, $\Gamma_{b.s}$) associated to the
  transition  about the ($ a_s$, $b_s$)
  cycles, respectively.  For the description of the above transitions
  we use  supersymmetrical versions of the Schottky  parameterization
   [12,13].   Then the above $(\Gamma_{a,s},\Gamma_{b,s})$
   are determined by $(3|2)$ complex parameters: two fixed
     supermanifold points $t_s^{(+)}=(u_s|\mu_s)$ and
     $t_s^{(-)}=(v_s|\nu_s)$,  as well as the multiplier $k_s $
     $(|k_s|<1,|\ arg k_s|\leq\pi)$.
     The replacement
   $\sqrt{k_s}\rightarrow-\sqrt{k_s}$ presents the supermodular
   transformation   that turns  the  genus-1 superspin
   structure $(l_{1s}=0 ,l_{2s}=1/2)$ into
     the  $(l_{1s}=0,l_{2s}=0)$
   superspin one.  To construct supermodular invariant amplitudes
     we   choose the set of  transition groups to
     be consistent with the above supermodular transformation.
     Therefore, we require that transition groups assigned to the above
     genus-1 superspin structures turn into each  other
     when $\sqrt{k_s}\rightarrow-\sqrt{k_s}$.

      If the odd parameters ($\mu_s,\nu_s$) are equal to zero,
     the  base transition
      group elements  can be chosen to be equal to
     ($\Gamma_{a,s}^{(o)}$, $\Gamma_{b,s}^{(o)}$) where
     \begin{equation}
     \Gamma_{a,s}^{(o)}(l_{1s})=\{z\rightarrow
     z ,\theta\rightarrow(-1)^{2l_{1s}}\theta\},\quad
     \Gamma_{b,s}(l_{2s})=
     \Gamma_{a,s}^{(o)}(1/2-l_{2,s})\Gamma_s^{(o)} .
     \end{equation}
      Here $\Gamma_s^{(o)}$=$\{z\rightarrow (a_sz+
     b_s)(c_sz+d_s)^{-1},\theta\rightarrow\theta(c_ sz+d_s)^{-1}\}$ and
     $a_sd_ s-b_sc_s=1$.  The $( a_s ,b_s ,c_s ,d_s)$ can be expressed in
     the terms of above $u_s$, $v_s$ and $k_s$, as well. The transitions
     (1) remain to be unchanged the  supermanifold points:
      $t_{o,s}^{(+)}=(u_s|0)$ and $t_{o,s}^{(-)}= (v_s|0)$.
     For arbitrary odd parameters ($\mu_s,\nu_s$) we  define the
    discussed base elements as \footnote{ The below transition groups
    differ from the ones given in refs. [2,14] except only
     for the ($l_{1s}=0 ,l_{2s}=1/2$) case.}
     \begin{equation}
     \Gamma_{ a,s}=\tilde\Gamma_s\Gamma_{ a ,s}^{(o)}(l_{1s})
     \tilde\Gamma_s^{-1},
    \qquad
    \Gamma_{b ,s}=\tilde\Gamma_s\Gamma_{b ,s}^{(o)}(l_{2s})
    \tilde\Gamma_s^{-1}=
    \Gamma_{ a,s}({1/2}-l_{2s})\Gamma_s\qquad
    \end{equation}
    where $\Gamma_s=\tilde\Gamma_s\Gamma_s^{(o)}\tilde\Gamma_s^{-1}$ and
    $\tilde\Gamma_s$ is a suitable transformation.
      It is convenient to require  that
   $t_{o,s}^{(-)}\rightarrow  t_s^{(-)}$ and
    $t_{o,s}^{(+)}\rightarrow t_s^{(+)}$ under the $\tilde\Gamma_s$
   mapping.   Then   $\tilde\Gamma_s$   is determined as
    \begin{eqnarray}
    \tilde\Gamma_s:\qquad z=z_s+\theta_s\tilde
    \varepsilon_s(z_s),\qquad
    \theta=\theta_s(1+\tilde\varepsilon_s\tilde
    \varepsilon_ s'/2)+
    \tilde\varepsilon_s(z_s);\nonumber\\
    \tilde\varepsilon_ s'=\partial_{z}\varepsilon_s(z),\qquad
    \tilde\varepsilon_s(z)=[\mu_s( z-v_s)-
    \nu_s( z-u_s](u_s-v_s)^{-1}.
    \end{eqnarray}

     Being  superconformal, all the above transitions preserve the
     spinor   derivative  $D(t)$ up to some  factor. For
     arbitrary  supersymmetrical   transformation
     $\Gamma=\{t\rightarrow t_\Gamma=(z_\Gamma(t)|\theta_
     \Gamma(t))\}$ this factor $Q_\Gamma(t)$ is
     \begin{equation}
    Q_\Gamma^{-1}(t)=D(t)\theta_\Gamma(t)\quad {\rm where}\quad
  D(t)=\theta\partial_z+\partial_\theta ;\quad
  D(t_\Gamma)= Q_\Gamma(t)D(t) .
    \end{equation}

  The  fundamental domain on the complex  z-plane is the region
 exterior to all the circles:$C_s^{(-)}=
 \{z :|Q_{\Gamma_{b ,s}}(t)|=1\}$
 and $C_s^{(+)}=\{z :|Q_{\Gamma_{b ,s}^{-1}}(t)=1\}$.
 We define the above region exterior (interior) to be the same as when all
 the odd parameters are reduced to zero.

 It is obvious from  eqs.  (1) - (3) that
 $\Gamma_{a,s}(l_{1s}=0)=I$,  $\Gamma_{a,s}^2(l_{1s})=I$ and
 $\Gamma_{a,s}(l_{1s}=1/2)$  is given by
 \begin{equation}
 \Gamma_{ a,s}(l_{1s}=1/2)=\{z\rightarrow z-2\theta\tilde
 \varepsilon_s(z)
 ,\quad\theta\rightarrow-\theta(1+2\tilde\varepsilon_s
 \tilde\varepsilon_s')+2\tilde\varepsilon_s(z)\}
 \end{equation}
 where $\tilde\varepsilon_s$  is defined by  eq.(3). Therefore, for
 $l_{1s}=1/2$  the cut $\tilde C_s$ appears on the considered  z-plane.
 One of its  endcut points  is placed  inside
 the $C_s^{(-)}$
 circle and the other  endcut point  is placed inside  the $C_s^{(+)}$ one.
  A  superconformal  p-form $F_p(t)$ changes under
 the ($\Gamma_{a,s} ,\Gamma_{b,s}$) transitions ($t\rightarrow
 t_{\Gamma_{a,s}}=t_s^a,t\rightarrow t_{\Gamma_{b,s}}=t_s^b$) as
 \begin{equation}
 F_p(t_s^a)=F_p^{(s)}(t)Q_{\Gamma_{a,s}}^p(t),\qquad
 F_p(t_s^b)=F_p(t)Q_{\Gamma_{b,s}}^p(t)
 \end{equation}
 where $F_p^{(s)}(t)$  is obtained by $2\pi$-twist of $F_p(t)$ about the
 $C_s^{(-)}$ circle.

 The  superspin structure $S_0=\bigcup_s(l_{1s}=0 ,l_{2s}=1/2)$  has been
 considered in refs. [5,6]. It  can also be obtained by the
 "naive" supersymmetrization  of the boson string [14]. The superspin
 structures $S(0,l_2)=\bigcup_s(l_{1s}=0,l_{2s})$ can be obtained from the
 above $S_0$ by the  $\sqrt k_s\rightarrow- \sqrt k_s$ replacement for every
 $\sqrt k_s$ associated with $l_{2s}=0$.  For the remained even superspin
 structures the Green functions are branched on z-plane that complicate
 their calculation. These  superspin structures $S_{br}$ are considered
 below.

 \section{Scalar supermultiplets}
 The having zero periods vacuum correlation function  of two identical
 scalar superfields can be written [2,6,13]
 in the terms of the holomorphic Green function $R(t,t')$,   its
 periods $J_r(t)$ and the period matrix $\omega=\{\omega_{sr}\}$.
 Owing to Bose statistics we have that
 $R(t,t')=R(t',t)$.   Also,
 \begin{equation}
 J_r(t_s^a)=J_r^{(s)}(t)+2\pi i\delta_{rs},\qquad J_r(t_s^b)=J_r(t)+2\pi i
 \omega_{sr}
 \end{equation}
 where ($t\rightarrow t_s^a,t\rightarrow t_s^b$)
 transitions are the same as in eq.(6). We normalize $R(t,t')$ by
 $R\rightarrow (z-z'-\theta\theta')^{-1}$ at $z\rightarrow z'$.  Apart
 from  unessential terms in $R(t,t')$ due to the scalar zero mode,
 both $R(t,t')$ and $J_s(t)$ are fully determined  by the equations:
 \begin{equation}
 R(t_s^a,t')=R^{(s)}(t,t'),\qquad R(t_s^b,t')=R(t,t')+J_s(t') .
 \end{equation}
 In ref. [6] we calculated $R$ in the explicit form only for
 the $S_0$ superspin structure.  Now we calculate it for every even
 superspin one.  Instead of $R(t,t')$ it is convenient for this aim   to
 have deal with the Green function $K(t,t')$ where $K(t,t')=
 D(t')R(t,t')$.
 Furthermore, we fix the above unessential terms in $R(t,t')$ up to an
 additive constant by the requirement:  $J_r\rightarrow 0$ at
 $z\rightarrow\infty$. Then $K(t,t')\rightarrow 0$ at
 $z\rightarrow\infty$ or $z'\rightarrow\infty$.

 The 1/2-differentials $\eta_s$ are defined as $2\pi
 i\eta_s(t)=D(t)J_s(t)$.  For discussed $S_{br}$ superspin structures all
 $\eta_s$ appear to be branched, if odd parameters are unequal to zero.
 To obtain  the normalization set for $\eta_s$
    we define for arbitrary 1/2-form $F_{1/2}$
    the integral $W_r(F_{1/2})$
    as\footnote{Throughout this paper the contour $C_r$ is defined to
 surround $(C_r^{(-)},C_r^{(+)})$ circles together with the
 $\tilde C_r$  cut, the $C_r$ contour being closed to
 $(C_r^{(-)},C_r^{(+)},\tilde C_r)$ above. Besides, $dt=d\theta dz$ and
 $\int d\theta\theta=1$.}
 \begin{equation} 2\pi iW_r(F_{1/2})\eta_r(t')=-\int
 \limits_{C_r}F_{1/2}(t)\frac{dt}{2\pi i}K(t,t')
 \equiv-E_r(F_{1/2},K)
 \end{equation}
 Using eqs.(8) and(9) one can obtain the above $W_r$ in the
 explicit form that is omitted here.Also, it can be proved that
 \begin{equation}
 2\pi iW_r(\eta_s)= \delta_{rs}\qquad{\rm{and}}\qquad
 W_s(K)=0
 \end{equation} where $W_s(K)$ is equal to
 $W_s(F_{1/2})$ at $F_{1/2}(t')=K(t,t')$.  Eqs.(10) are proved by
 moving to infinity the integration contour in
 $\sum\nolimits_{r=1}^{n}E_r(F_{1/2},K)$ for $F_{1/2}$ to be equal
 to $\eta$ or $K$.  Besides, the linear independence of the
 different $\eta_s$  is taken into account.  Eqs.(10) turn out to
 be useful for the calculation of the Green function $K(t,t')$.

 If all odd parameters are equal to zero, $K(t,t')$ reduces to
 $K^{(o)}(t,t')=K_b(z,z')+\theta\Phi(z,z')(z-z')^{-1}$.
 The $K_b(z,z')$ is given by series over Schottky
 group elements [2,13]. The like series determining
 $\Phi(z-z')^{-1}$ may be divergent for the considered $S_{br}$
 structures, but in any case $\Phi$ can be written as
 \begin{eqnarray}
 \Phi(z,z')=\left({\prod}^\prime
 \frac{[\phi_\Gamma (z)\phi_\Gamma(z')]^{1/2}} {[z-g_\Gamma(z')][z'
 -g_\Gamma(\infty)]}
 \right)
 \frac{\Theta[l_1,l_2](J|\omega)}{\Theta[l_1,l_2](0|\omega)}
 \nonumber\\
 {\rm{where}}\quad\phi_\Gamma(z)=[z-g_\Gamma(z)][z-g_\Gamma
 (\infty)].
 \end{eqnarray}
 Here  $\Theta$  is the theta function. The symbol $J$ denotes
 set of $\left(J_s^{(o)}(z)-\right.$\\$\left. J_s^{(o)}(z')\right)$
 functions.  To every $\Gamma$
 the mapping $z\rightarrow g_\Gamma(z)$
 is assigned.
 The product ${\prod}^\prime$ includes all Schottky group elements
 $\Gamma$ except only for $\Gamma=I$.

 Odd parameters being arbitrary, to calculate Green functions  $K$
 for the discussed $S_{br}$ superspin structures we construct
 the set $\{K_s^{(o)}\}$
 of "master" Green functions.
 For every $K_s^{(o)}(t,t')$ its transition group
 elements associated with rounds about the $(a_s,b_s)$ cycles are
 constructed to be the same as for the truly Green function
 $K(t,t')$.
 However, the transition group elements associated with rounds
 of $K_s$  about all the other cycles may differ from those
 assigned to the above Green function $K(t,t')$. At last,
 $K_s(t,t')\rightarrow0$ at $z\rightarrow\infty$ or
 $z'\rightarrow\infty$.

 To calculate $K(t,t')$ we start with the following
 relations:
 \begin{equation}
 K(t,t')=K_s^{(o)}(t,t')+\sum_{r=1}^{n}
 \int\limits_{C_r}K_s^{(o)}(t,t'')
 \frac{dt''} {2\pi i}K(t'',t')\quad{\rm{for}}\; s=1,2...n.
 \end{equation}
 Eq.(12) can be verified by moving  of the integration contour
 $\bigcup C_r$ to infinity. Then the nonzero contribution
 originates from the poles at $z''=z$ and $z''=z'$.
 In the sum over $r$  there is no the term corresponding to $r=s$.
 Indeed, this term can be written as $2\pi iW_s(K_s^{(o)})
 \eta_s(t')$ where
 $W_s$ is defined by eq.(9) for $F_{1/2}(t')=K_s^{(o)}(t,t')$.
 So, this term vanishes owing to eq.(10).
 For $z\in\bigcup C_s$ we define the set $\tilde K=
 \{K_s(t,t')\}$ by the
 relations: $K(t,t')=K_s(t,t')$, if $z\in C_s$. The
 above set can be  calculated from eqs.(12).
 As soon as the $C_s$ contours can be moved, all the
 above $K_s$, in fact,  determine the same function
 $K(t,t')$.  So, a choice of either $K_s$ to fit the discussed
 $K$ is only a matter of convenience.\footnote{We omit here
 the rigorous proof of this statement.} The eqs.  (12) for
 the above $\tilde K$ set can be also written as
 \begin{equation}
 \tilde K=\tilde K^{(o)}+\hat K^{(o)}\tilde K
 \end{equation}
 where $\tilde K^{(o)}=\{K_s^{(o)}\}$.  The matrix  $\hat
 K^{(o)}=\{\hat K_{sr}^{(o)}\}$ is the integral operator, its kernel
 being $K_{sr}^{(o)}$.  Here  $K_{sr}^{(o)}$  is defined by the relations:
 $K_{sr}^{(o)}(t,t')=K_s^{(o)}(t,t')$,
 if $z\in C_s$, $z'\in C_r$  and
 $s\neq r$;   $K_{ss}^{(o)}=0$.

 Below we use two sets of master Green functions $K_s^{(o)}$.
 Firstly, we construct $K_s^{(o)}$ as
 \begin{equation}
 K_s^{(o)}(t,t')=K^{(o)}(t_s,t_s')D(t')\theta_s'+
 \tilde\varepsilon_s'\theta_s'\Phi(\infty,z_s')
 \end{equation}
 where $t_s$ is calculated in the terms of $t$  from
 eqs.(3).  Then $K_s$  can be calculated from eq.(13)
 by the iteration procedure, every posterior iteration
 being one more power in odd parameters than a previous
 one. Therefore,  $K_s$ appears to be a series containing a finite
 number of terms.

 The second set   we construct  in the terms of
 the genus-1 and genus-2 Green functions. The genus-1 Green function we
  assign to every handle except only for handles associated with the
  odd genus-1 superspin structure:  $l_{1s}=l_{2s}=1/2$. The number
 of the latter handles is even for even genus-n superspin structures
 and   we group them  into pairs.
 Then to every pair we assign the genus-2 Green function that is
 calculated from  eq.(13), where $K_s^{(o)}$ are defined by
 eq.(14). The genus-1 Green functions  are given by eq.(14).
 We denote the considered set as $\tilde K_0+\tilde \Xi$ where
 $\tilde K_0$  is calculated at all the odd parameters to be equal
 to zero and $\tilde\Xi$ is  proportional to odd Schottky parameters.
   Then  eq.(13) can be turn into the following one:
 \begin{equation}
 \tilde K=\tilde K_0^{(o)}+\tilde\Xi+
 \hat
 K_0^{(o)}\tilde\Xi+
 (\hat\Xi+ \hat K_0^{(o)}\hat\Xi)\tilde K .
 \end{equation}
 In eq.(15) the operator  $\hat K_0^{(o)}$ ($\hat\Xi$) is related with
  $\tilde K_0^{(o)}$ ( $\tilde\Xi$ ) just as $\hat K^{(o)}$
 is related with $\tilde K^{(o)}$  in eq.(13). Besides, $\tilde
 K_0^{(o)}= \{K_{0s}^{(o)}\}$ where $K_{0s}^{(o)}$ are expressed in the
 terms of the reduced Green function
 $K^{(o)}$: $K_{0s}^{(o)}(t,t')=K^{(o)}(t,t')$ at $z\in C_s$.

When $K$ is determined one easy calculate the Green function
$R(t,t')$ up to unessential additive constant. If
 $z$ is situated near the $C_s$ contour, it is convenient to write its
 periods $J_s(z)$ as
 \begin{equation}
 J_s=J_s^{(o)}+\sum_p\hat K_{sp}^{(o)}\hat
 K_{ps}J_s^{(o)}
 \end{equation}
 where $J_s^{(o)}$ is the period of $R_s^{(o)}$ corresponding
 to $2\pi$-twist about the $b_s$-cycle.  The set
 $\hat K$ of the  integral operators
 $\hat K_{ps}$  is calculated from the equations:
 \begin{equation}
 \hat K=\hat K^{(o)}+\hat K^{(o)}\hat K.
 \end{equation}
 For $z$ to be situated near the $C_r$ contour $( r\neq s )$ one
 can write:
 \begin{equation}
 J_s=\hat K_{rs}^{(o)}J_s^{(o)}+\sum_p\hat K_{rp}^{(o)}\hat
 K_{ps}J_s^{(o)} .
 \end{equation}

 The period matrix $\omega$ turns out to be
 \begin{eqnarray}
 \omega_{sr}=\eta_r^{(o)}\hat J_s^{(o)}+\sum_{p\neq r}
 \eta_r^{(o)}\hat K_{ps}J_s^{(o)}\qquad{\rm{for}}\quad r\neq s
 \nonumber\\ \qquad {\rm{and}}\qquad
 \omega_{ss}=\omega_{ss}^{(o)}+\sum_{p\neq s}
 \eta_s^{(o)}\hat K_{ps}J_s^{(o)}.
 \end{eqnarray}
 In eq. (19)  $\omega_{ss}^{(o)}$ is
 the $(ss)$ element of the period matrix associated with
 $J_s^{(o)}$  The  integral operator $\hat J_s^{(o)}$
 is defined for $z\in C_s$,
 its kernel being $J_s^{(0)}(t)$.
 One can verify from eqs.(19)   that  the period matrix
  depends on the superspin structure owing to the terms
  proportional to the odd parameters. It seems natural since
 these terms appear because  fermions mix  bosons under
 the supersymmetry  transformations.

 \section{Ghost supermultiplets}
 In the considered scheme [5,6] both the supermoduli volume
 form and zero mode contributions to the ghost determinant are
 counted by using of a suitable ghost vacuum correlation function
 $G_{gh}(t,t')$. The discussed $G_{gh}$ can be expressed
 [5,6] in the terms of the Green function $G(t,t')$ and
 superconformal 3/2-forms $\chi_N(t')$, all they being
 fully determined by the relations:
 \begin{eqnarray}
 G(t_r^a,t')=Q_{\Gamma_{a,r}}^{-2}(t)\left(G^{(r)}(t,t')+
 \sum_{ N_r}
 Y_{a,N_r}(t)\chi_{N_r}(t')\right)\nonumber\\
 G(t_r^b,t')=Q_{\Gamma_{b,r}}^{-2}(t)\left(G(t,t')+
 \sum_{N_r}Y_{b,N_r}(t) \chi_{N_r}(t')\right)\nonumber\\
 G(t,t_r'^a)=Q_{\Gamma_{a,r}}^{-2}(t)G^{(r)}(t,t'),\qquad
 G(t,t_r'^b)=Q_{\Gamma_{b,r}}^{-2}(t)G(t,t')
 \end{eqnarray}
 where $N_r=k_r,u_r,v_r, \mu_r$ or $\nu_r$. Furthermore,
 $Q_{\Gamma_r}^{-2}Y_{q,N_r}=
 \partial_{N_r} g_r^q+\gamma_r^q\partial_{N_r}\gamma_r^q$
 with $q=a,b$. The above $Y_{q,N}$ are power-2 polynomial
 in $(z,\theta)$.
 For $l_{2r}=1/2$ the functions $Y_{b,N_r}$
 have been calculated  in ref. [6].
 Both $t_r^a=(g_s^a|\gamma_r^a)$ and $t_r^b=(g_r^b|\gamma_r^b))$
 in eqs.(20) are the same as in
 eqs.(6). The relations (20) generalize the results of refs.
 [5,6] to arbitrary superspin structures.

 For an arbitrary 3/2-form $F_{3/2}$ we define  the integral
 $H_{N_r}(F_{3/2})$ by the relation:
 \begin{equation}
 2\pi i\sum_{N_r}H_{N_r}(F_{3/2})\chi_{N_r}(t')=
 -\int_{C_r}F_{3/2}(t)\frac{dt}{2\pi i}G(t,t') .
 \end{equation}
 Using eqs.(20) and (21) one can calculate $H_N$ in
 the explicit form, but we omit it here.Also, it can be proved that
 \begin{equation}
 2\pi iH_{N_s}(\chi_{N_r})=\delta_{N_s,N_r}\qquad{\rm{and}}\qquad
 H_{N_s}(G)=0
 \end{equation}
 where $ H_{N_s}(G)$ is equal to
$H_{N_s}(F_{3/2})$ calculated at $F_{3/2}(t')=G(t,t')$.
Eqs.  (21) - (22) are similar to eqs.  (9) - (10) for scalar
 supermultiplets.

 If all odd parameters are equal to zero, the Green function
 $G(t,t')$  is reduced to the Green function
 $G^{(o)}(t,t')=G_b(z,z')\theta'+\theta G_f(z,z')$.  Being
 independent of the spin structure, $G_b$  can be  obtained  from the
 ghost Green function given in refs.
 [5,6].  The discussed $G_f$ can be calculated in the terms
 of Green functions $G_{\{\sigma\}}(t,t')$ defined as
 \begin{equation}
 G_{\{\sigma\}}(t,t')= \sum_\Gamma\frac{\exp\pi
 i[\Omega_\Gamma(\{\sigma_s\})+\sum_s2l_{1s}
 \sigma_s(J_s(z)-J_s(z'))]}
 {[z-g_\Gamma(z')]Q_\Gamma^3(z')}
 \end{equation}
 where $\sigma_s=\pm1$. So, $G_{\{\sigma\}}$ depends on a choice
 of the $\{\sigma_s\}$ set. In eq.
 (23) the summation performs over all Schottky group elements
 $\Gamma$, the base ones being $\Gamma_s$. The value
 $\Omega_\Gamma(\{\sigma_s\})$ is
 \begin{equation}
 \Omega_\Gamma(\{\sigma_s\})=-\sum_{s,r}2l_{1s}
 \sigma_s\omega_{sr}n_r(\Gamma)+
 \sum_r(2l_{2r}-1)n_r(\Gamma)
 \end{equation}
 where $n_r(\Gamma)$ is the
 number of times that the $\Gamma_r$ generators  are present
 in $\Gamma$
(for its inverse $n_r(\Gamma)$ is defined to be negative ).

 The changes of $G_{\{\sigma\}}$ under the $(t\rightarrow t_r^b)$
 transitions are given by
 \begin{eqnarray}
 G_{\{\sigma\}}(t_r^b,t')=Q_{\Gamma_{b,r}}^{-2}(t)
 \left(G_{\{\sigma\}}(t,t')
 +\sum_{N_r}\tilde Y_{b,N_r}(t)
 \chi_{\{\sigma\},{N_r}}(t')\right)\nonumber\\
 {\rm{where}}\qquad\tilde Y_{b,N_r}(t)=\exp[\pi
 i\sum\nolimits_s2l_{1s}\sigma_sJ_s(t)]Y_{b,N_r}(t)
 \end{eqnarray}
 and  $\chi_{\{\sigma\},{N_r}}(t')$  are
  3/2-forms.  The discussed $G_f(t,t')$ turns out to be
  \begin{equation}
 G_f(t,t')=G_{\{\sigma\}}(t,t')-\sum_N\tilde H_N(t)\chi_N(t')
 \end{equation}
 where $\tilde H_N(t)$ is $H_N(F_{3/2})$ calculated at
 $F_{3/2}(t')= G_{\{\sigma\}}(t,t')$. The index $N$ labels
 the even and odd Schottky parameters. As soon as
 eqs.(20) determine $G_f(t,t')$ in the unique way,
 $G_f(t,t')$ given by
 eq.(26) is independent of $\{\sigma\}$. Also, from eqs.(20)
 and (26) it  follows that
 \begin{equation}
 \chi_{\{\sigma\},N}=\sum_{N'}M_{N,N'}(\sigma)\chi_{N'}\quad
 {\rm{where}}\quad M_{N,N'}(\sigma)= H_{N'}(\chi_{\{\sigma\},N}) .
 \end{equation}

 For arbitrary odd parameters  the discussed $G(t,t')$ is
 calculated by the same method as  $K(t,t')$
 considered in the previous section.

 \section{Multiloop superstring amplitudes}
 Using both the above Green
 functions and eqs.(22) for $\chi_N$ we calculate the partition
 functions from the equations derived in refs. [5,6].
 Below we fix $(3|2)$ Schottky parameters
 $(u_1,v_1,u_2,\mu_1,\nu_1)$ to be the same for all
 the genus-n supermanifolds, the rest of $(3n-3|2n-2)$
 Schottky ones being chosen as  moduli.
 In the closed, oriented superstring theory the m-leg, n-loop
 amplitudes $A_n^m$ are given by
 \begin{eqnarray}
 A_n^m=g^n\int\sum_{(L,L')}\{\det
 2\pi[\omega(L)-\overline{\omega(L')}]\}^{-5}Z_L(N)\overline
 {Z_{L'}(N)}B_{LL'}\times\nonumber\\
 \prod_{N'} dN'
 d\overline{N'}\prod_{p=1}^mdt_pd\overline{t_p}
 \end{eqnarray}
 where the line over denotes the complex conjugation, $g$
 is a coupling constant and $N'$ label those
 Schottky parameters $N$ that are chosen to be moduli.
 The summation performs over all the superspin
 structures $L$ and $L'$ of the right and left fields.  Also,
 $B_{LL'}=B_{LL'}(\{t_p\},\{\overline t_p\})$  are the vertex products
 integrated over fields.  Using the boson emission vertices [13] and
 the results obtained in Sec.3 of the presented paper one can calculate
 the discussed $B_{LL'}$ for the boson emission amplitudes.
 The fermion emission amplitudes need a further study.

 In eq.(28) the factors $Z_L(N)$ are holomorphic in $N$.
  Therefore, Belavin-Knizhnik theorem [8] is correct for every
 term in eq.(28), but its naive generalization to $A_n^m$ is not true
 because the period matrix $\omega=\omega(L)$ given by eq.(19)
 depends on $L$.  The discussed   $Z_L(N)$  factors  turn
 out to be
 \begin{equation} Z_L= Z_L^{(o)}\tilde Z_L
 [(u_1-u_2)(v_1-u_2)-\mu_1\mu_2/2-\nu_1\mu_2/2]
 \prod_{r=1}^n(u_r-v_r-\mu_r\nu_r)^{-1} .
 \end{equation}
 The factor $Z_L^{(o)}$ in eq.(29) is calculated at
 all the odd  parameters  equal to zero. The contribution
 of the odd parameters is counted by the factor  $\tilde Z_L$.
 In the equation for $Z_L^{(o)}$ a lot of
 terms  vanishes  because these terms  can be written as
 $\exp[\pi i\Omega_\Gamma](z-g_\Gamma(z))^{-1}
 Q_\Gamma^{-3}(z)\partial_NJ_r(z)$ integrated along
 $\bigcup C_s$.  The result is
 \begin{eqnarray}
 Z_L^{(o)}=\frac{\Theta^5[l_1,l_2](0|\omega)\exp[-\pi
 i\sum\nolimits_{j,r}
 l_{1j}l_{1r}\omega_{jr}]}{\Theta^5[\{0\},\{1/2\}]
 (0|\omega)\sqrt{\det M(\sigma)M(-\sigma)}}
  \prod_{j=1}^n\frac{1+\sqrt
 k_j\lambda_j}{k_j^{3/2}\lambda_j}\times\nonumber\\
 \prod_{(k)}(1+\sqrt  k)^{-2}\prod_{m=1}^\infty
 \frac{(1-k^m)^{-8}(1-k^{m-1/2})^8 (1-k^{m+1/2})^2}
 {[1-\Lambda_\Gamma (\sigma)k^{m+1/2}][1-
 \Lambda_\Gamma(-\sigma)k^{m+1/2}]}
 \end{eqnarray}
 where $\lambda_j= (-1)^{(1-2l_{2j})\pi i}$ and
 $\Lambda_\Gamma(\sigma)=\exp[\pi i\Omega_
 \Gamma(\{\sigma_s\})]$, the above
 $\Omega_\Gamma(\{\sigma_s\})$ being defined by eq.(24). The
 matrix $M(\sigma)$ is defined by eq.(27) and $\Theta$ is
 the theta function. The $\theta$ in the denominator associates with
 the $S_0$ spin structure. The product over $(k)$ is taken over
 all multipliers of the Schottky group,  which are not powers of
 other ones.  In fact, eq.(30) does not depend on a choice of
 the $\{\sigma\}$ set because Green function $G_f(t,t')$
 given by eq.(26) is independent of $\{\sigma\}$. To avoid
 misunderstanding we note one more that the right side of eq.(30) is
 calculated at all odd parameters  to be equal to zero.

 To calculate the factor  $\tilde Z_L$  in eq.(29) we use that both
  $\partial_NK(t,t')$ and $\partial_NG(t,t')$ can be written as
 the integral $H_N(F_{3/2})$
 defined by eq.(21) with a suitable $F_{3/2}$. For $\partial_NK$ the
 $F_{3/2}$ form appears to be by-product of $K$  and its
 derivatives in respect to $z$ or $\theta$.For $\partial_NG$ the
 discussed $F_{3/2}$ form includes, besides, by-products of
 3/2-differentials and the factors that are power-2 polynomial
 in  ($z,\theta)$. For $n=2$  we found that
 \begin{equation}
 \tilde Z_L=trace[-5\hat K^{(o)}\hat K^{(o)}+
 \hat G^{(o)}\hat G^{(o)}]/4
 \end{equation}
 where the operator $\hat G^{(o)}$
 is associated with the ghost supermultiplets in the same way as
 the integral operator $\hat K^{(o)}$
 associated with the scalar ones. The operator
  $\hat K^{(o)}$  is the same as in  eq. (13), where
  $\tilde K^{(o)}$  is defined by eq.(14).
 For $n>2$ the result is
 \begin{equation}
 \tilde Z_L=trace[5\ln(1-
 \hat \Xi-\hat \Xi\hat K_o^{(o)})-\ln(1-\hat \Psi-
 \hat \Psi\hat
 G_o^{(o)})]+\sum_p\tilde Z^{(2)}(s_p,r_p) .
 \end{equation}
 In eq.(32) the operators $\hat G_0^{(o)}$ and $\hat\Psi$ are
 associated with the ghost supermultiplets in the same way as
 $\hat K_0^{(o)}$ and  $\hat\Xi$ associated with the scalar ones,
  both $\hat K_0^{(o)}$ and $\hat\Xi$ being the same as in
 eq (15).Every $Z^{(2)}(s_p,r_p)$ denotes the genus-2
 contribution (31)
 due to  the pair of handles $(s_p,r_p)$, every handle being associated
 with the odd genus-1 superspin structure. For
 all the above handles to be grouped into pairs,
 the summation  performs over all the above pairs. The considered sum,
   as well as both $\hat\Xi$ and $\hat\Psi$, depends on a choice of
   the dividing of the considered handles into pairs, but  $\tilde
   Z_L$ is independent of the above choice.

 \section{The problem of divergences}
 In eq.(28) the integration region  over $N'$  is determined by the
 supermodular invariance. Without a loss of generality one can
 exclude from this region those domains where
 some of the Schottky group multipliers $k$ are near to
 one: $k\approx 1$. Indeed, modulo of supermodular
transformations, these  domains are equivalent to those  where some
of $k_j$ are small: $k_j\approx 0$.
At $k_j\rightarrow 0$
we see from eq.(30)  that   $Z_L\sim k_j^{-1}$
 for $l_{1j}=1/2$ and   $Z_L\sim k_j^{-3/2}$
 for  $l_{1j}=0$.
 However, in the sum(28) over $(L)$ the above singularity
 $k_j^{-3/2}$ is reduced to $k_j^{-1}$. Besides, we have
 the factor $(\ln|k|)^{-5}$ due to
 $\det[\omega(L)-\overline{\omega(L')}]^{-5}$
 in eq.(28).
 As the result, the integral(28) appears to be  finite at
 $k_j\rightarrow 0$.

 Nevertheless, the problem of the finiteness of the considered
 theory needs a further study. It follows from eq.(29)
 that, beside the above singularities at $k_j\rightarrow 0$, every
 $Z_L$ has also the singularities at
 $u_j-v_j\rightarrow 0$. One can interpret the above limit as
 the moving of the $j$-handle away from the others.
 The contribution to $A_n^m$ from the region where
 $u_j-v_j\rightarrow 0$ appears to be proportional to
 \begin{equation}
 \int\frac{d(Reu_j)d(Rev_j) d(Imu_j)d(Imv_j) d\mu_jd\nu_j
 d\overline\mu_jd\overline\nu_j}{|u_j-
 v_j-\mu_j\nu_j|^2}[Z_{n-1}A_1^m+(u_j-v_j)^2B]
 \end{equation}
 where $B$ is finite at $u_j-v_j\rightarrow 0$ and $Z_{n-1}$ is
 the genus-(n-1) partition function. One can see that
 the integral (33)
 has   uncertainty, if  $Z_{n-1}\neq 0$.
 The uncertainties of the same type arise also from
 the other regions, which correspond
 to the moving of the handles away from each other.
 The equality $Z_n=0$  has been argued in ref.
 [9]
 under the assumption that the discussed
 theory possesses the ten-dimensional supersymmetry,
 as well as the two-dimensional one. So, if the above assumption is true,
 the closed, oriented string appears to be free from
 the divergences. However, we did not verify the discussed assumption
 because of the mathematical complexity of this verification.

\end{document}